\documentclass[11pt,letterpaper]{article}
\usepackage{systeme} 
\usepackage[utf8]{inputenc}
\usepackage{multirow} 
\usepackage{booktabs} 
\usepackage[english]{babel}
\usepackage{amsmath}
\usepackage{amsfonts}
\usepackage{amssymb}
\usepackage{graphicx}
\usepackage[small,bf]{caption} 
\usepackage[left=3cm,right=3cm,top=2cm,bottom=3cm]{geometry}

\author{Jorge Pinochet}
\title{\textbf{The little robot, black holes, and spaghettification}}
\begin{document}

\author{Jorge Pinochet$^{*}$\\ \\
 \small{$^{*}$\textit{Departamento de Física, Universidad Metropolitana de Ciencias de la Educación,}}\\
 \small{\textit{Av. José Pedro Alessandri 774, Ñuñoa, Santiago, Chile.}}\\
 \small{e-mail: jorge.pinochet@umce.cl}\\}

\date{}
\maketitle

\begin{center}\rule{0.9\textwidth}{0.1mm} \end{center}
\begin{abstract}
\noindent The tidal forces generated by a black hole can be so powerful that they cause unlimited stretching, known as spaghettification. A detailed analysis of this phenomenon requires the use of Einstein's theory of general relativity. The aim of this paper is to offer an up-to-date and accessible analysis of spaghettification, in which the complex mathematics of Einstein's theory are replaced with the simpler and more intuitive concepts of Newtonian gravity. The article can serve as educational material for undergraduate modern physics or astronomy courses. \\ \\


\begin{center}\rule{0.9\textwidth}{0.1mm} \end{center}
\end{abstract}

\maketitle

\section{Introduction}
One of the most extreme effects caused by the gravity of a black hole is \textit{spaghettification}, a phenomenon popularised by Stephen Hawking in his best seller \textit{A Brief History of Time} [1]. To illustrate this effect, Hawking imagines an unfortunate astronaut falling into a black hole. As he descends, the astronaut begins to feel his body gradually stretch. This sensation becomes more and more uncomfortable until, when he is close enough, the astronaut undergoes an unlimited stretching that turns him into a long, thin line of subatomic particles; that is to say, he is "stretched like spaghetti" [1,2]. Although a detailed description of this phenomenon requires the use of Einstein's theory of general relativity, which has a well-deserved reputation for being mathematically very complex, the underlying explanation is simple and resides in the tidal forces that are a side effect of gravity and which can be described within the framework of Newtonian physics.\\

The aim of this work is to analyse spaghettification using the accessible and intuitive ideas of Newton's law of gravitation. Although an interesting article previously published in this journal has also addressed spaghettification [3], the analysis that we will develop here is broader and more detailed. Instead of resorting to the raw image of a spaghettified astronaut, we will describe spaghettification by means of a more impersonal image, in which we imagine a Little Robot, LR, that undertakes a one-way trip into a black hole. For simplicity, we will focus on a \textit{static black hole}, whose mathematical description depends only on its mass.

\section{Black holes and the Schwarzschild radius}
A black hole is a region of spacetime bounded by a closed spherical surface called the \textit{horizon}. Within the horizon, there is such a high concentration of matter that to escape its gravity requires a speed greater than that of light in vacuum, $c = 3\times 10^{8} m\cdot s^{-1}$. Since $c$ is the maximum speed allowed by relativistic physics, matter and energy can only cross the horizon from the outside to the inside, and never in the opposite direction. In the framework of Newtonian gravity, the horizon of a static black hole can be intuitively visualised as a spherical surface whose radius only depends on the mass concentrated inside\footnote{It should be noted that this image has only pedagogical value, since in general relativity, $R_{S}$ is a coordinate and does not represent the physical distance between the horizon and the central singularity.} [4]:

\begin{equation} 
R_{S} = \frac{2GM}{c^{2}}.
\end{equation}

For historical reasons, $R_{S}$ is called the \textit{Schwarzschild radius}, where $G = 6.67 \times 10^{-11} N\cdot m^{2} \cdot kg^{-2}$ is the constant of gravitation and $M$ is the mass of the black hole. A useful way to write this equation is as a function of the solar mass, $M_{\odot} = 1.99 \times 10^{30} kg$:

\begin{equation} 
R_{S} (m) \cong 3\times 10^{3} \left( \frac{M}{M_{\odot}} \right). 
\end{equation}

where $R_{S}$ is expressed in metres ($m$). Thus, an object whose mass is equal to that of the Sun will have a Schwarzschild radius of $3 \times 10^{3} m = 3 km$. According to general relativity, if an object is compressed within its horizon, a gravitational collapse occurs that no force can stop, which culminates when all the mass is concentrated at a mathematical point of zero size and infinite gravity called \textit{singularity} (see Fig. 1). Although the horizon is not a material surface, it can be imagined as a unidirectional membrane that only allows for the inward flow of matter and energy [1].\\

\begin{figure}[h]
  \centering
    \includegraphics[width=0.3\textwidth]{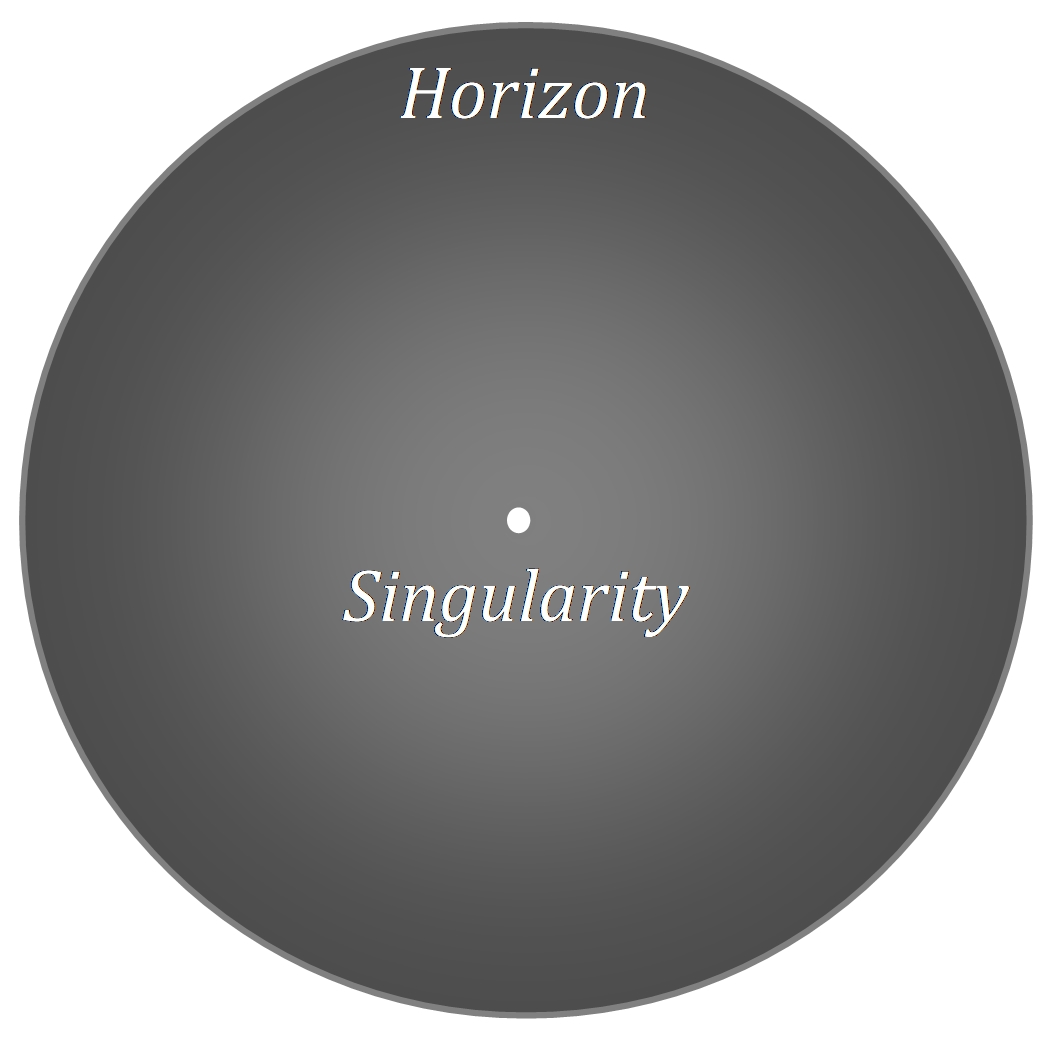}
  \caption{Structure of a static black hole.}
\end{figure}

Is every object that approaches the horizon spaghettified? Is it necessary to cross the horizon inward for spaghettification to occur? How do tidal forces vary as a function of $M$ and distance to the singularity? In the following sections, we will try to answer these and other questions.

\section{Universal gravitation and tidal forces}

According to Newtonian physics, the magnitude of the gravitational force between a particle of mass $m$ and a spherical body of uniform mass $M$ is [5]:

\begin{equation} 
F = \frac{GMm}{r^{2}},
\end{equation}

where $r$ is the distance between the particle and the centre of the spherical body. This equation shows that the force of gravity is not constant but varies from point to point. This variation is responsible for the tidal forces illustrated in Fig. 2, where four particles of mass $m$, designated as \textit{A}, \textit{B}, \textit{C}, \textit{D}, are affected by the gravity of a massive spherical body of mass $M \gg m$. Since $m$ is small, we can neglect the force of gravity between the four particles versus the force exerted by the sphere.\\

\begin{figure}[h]
  \centering
    \includegraphics[width=0.6\textwidth]{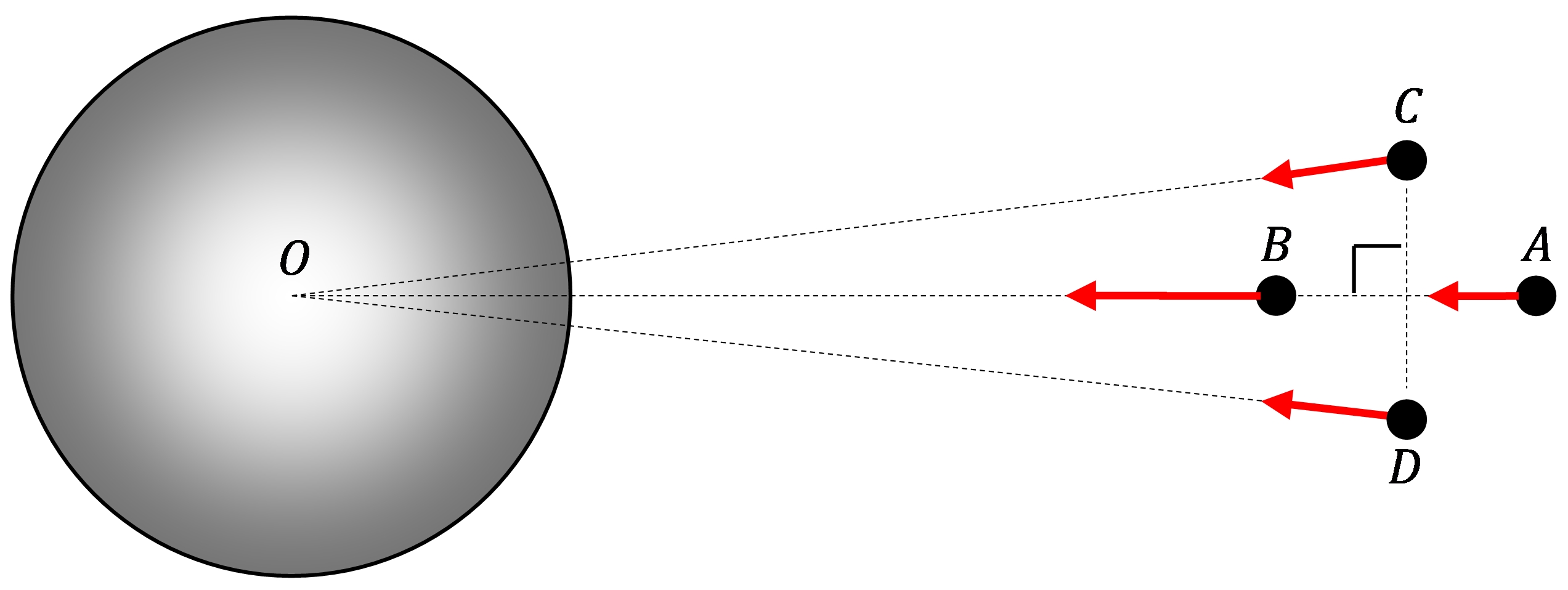}
  \caption{The gravity exerted by $M$ causes \textit{A} and \textit{B} to separate radially, and \textit{C} and \textit{D} to approach transversely. The line joining \textit{C} and \textit{D} is perpendicular to the radial direction through \textit{A} and \textit{B}.}
\end{figure}

It can be seen that \textit{A} and \textit{B}, which are positioned along the radial direction, experience different forces of gravity; the force on \textit{B} is greater than on \textit{A}, which causes them to separate further, and we call this the \textit{radial tidal force}. On the other hand, \textit{C} and \textit{D}, which are positioned in the transverse direction, experience gravitational forces that have the same magnitude but different directions; both of these forces point towards the centre O of the sphere, which causes the particles to approach each other, and we call this the \textit{transverse tidal force}. If we extend these ideas to any solid object made up of many particles, the conclusion is that the object will experience radial stretching and transverse compression as it approaches a massive spherical body. The difference in the effects caused by the radial and transverse tidal forces means that the signs are different: the radial force is negative, while the transverse force is positive. However, to simplify the analysis, we will only consider the absolute values of these forces in the following.\\

Fig. 3 illustrates the action of the radial forces generated by the sphere on \textit{A} and \textit{B}. Let $\Delta r$ be the separation between the particles, and let $r \gg \Delta r$ be the radial distance between the centre of the sphere and the midpoint between \textit{A} and \textit{B}. According to Eq. (3), the difference in the magnitude of the forces between the two particles in the radial direction is:

\begin{equation} 
\Delta F_{R} = F_{B} - F_{A} = \frac{GMm}{(r - \Delta r/2)^{2}} - \frac{GMm}{(r + \Delta r/2)^{2}}.
\end{equation}

After some algebra, we get:

\begin{equation} 
\Delta F_{R} = \frac{2GMmr\Delta r}{\left[ ( r^{2} - (\Delta r/2)^{2} \right]^{2}} = \frac{2GMm\Delta r}{r^{3} \left( 1- (\frac{\Delta r}{2r})^{2} \right)^{2}}.
\end{equation}

Since $r \gg \Delta r$, we can set $(\Delta r/2r)^{2} \cong 0$, giving:

\begin{equation} 
\Delta F_{R} = \frac{2GMm\Delta r}{r^{3}}.
\end{equation}

\begin{figure}[h]
  \centering
    \includegraphics[width=0.7\textwidth]{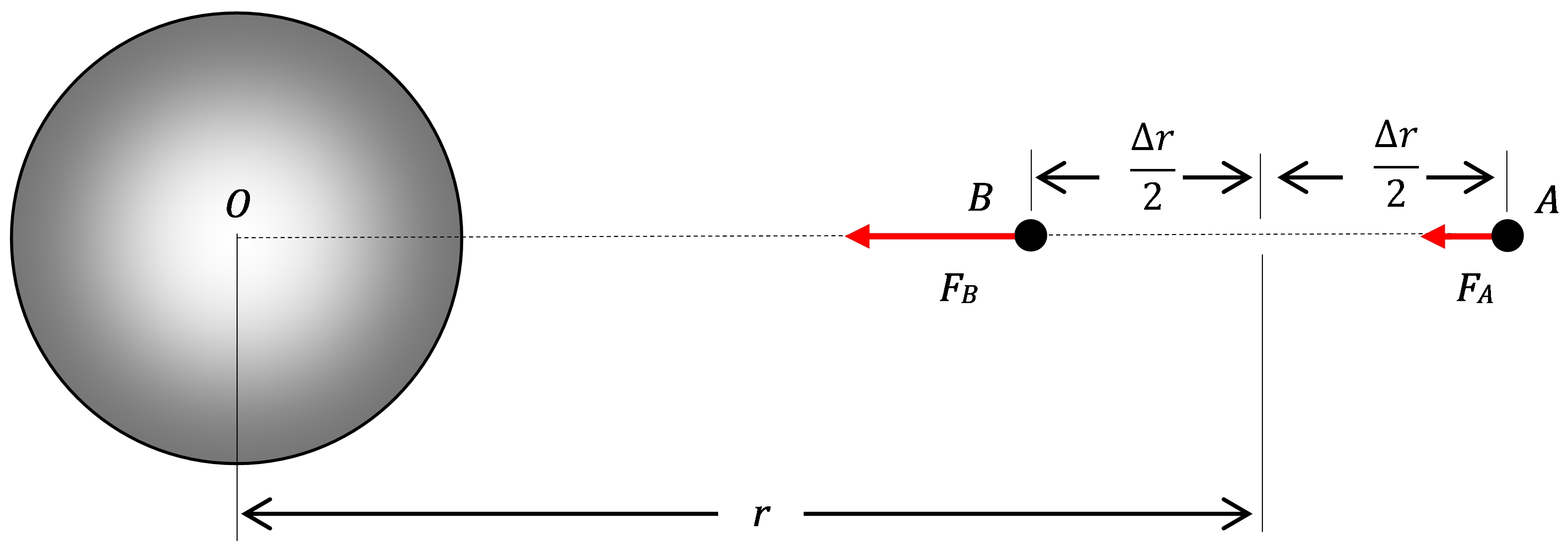}
  \caption{Radial tidal forces on \textit{A} and \textit{B}, where $F_{B} > F_{A}$.}
\end{figure}

Let us now calculate the tidal force in the transverse direction. Fig. 4 illustrates this situation, where $\Delta R$ is the separation between \textit{C} and \textit{D}. The particles are separated by a distance $r$ from the centre of the sphere. As shown in the figure, we can calculate the difference in the magnitudes of the forces between \textit{C} and \textit{D} in the transverse direction as:

\begin{equation} 
\Delta F_{T} = F_{C,T} - F_{D,T} = F_{C} sin\theta - F_{D} sin\theta.
\end{equation}

The gravitational forces on \textit{C} and \textit{D} are

\begin{equation} 
F_{C} = \frac{GMm}{r^{2}},\     F_{D} = -\frac{GMm}{r^{2}},
\end{equation}

and since $sin \theta = (\Delta R/2)/r$, we obtain:

\begin{equation} 
\Delta F_{T} = \frac{GMm\Delta R}{r^{3}}.
\end{equation}

\begin{figure}[h]
  \centering
    \includegraphics[width=0.7\textwidth]{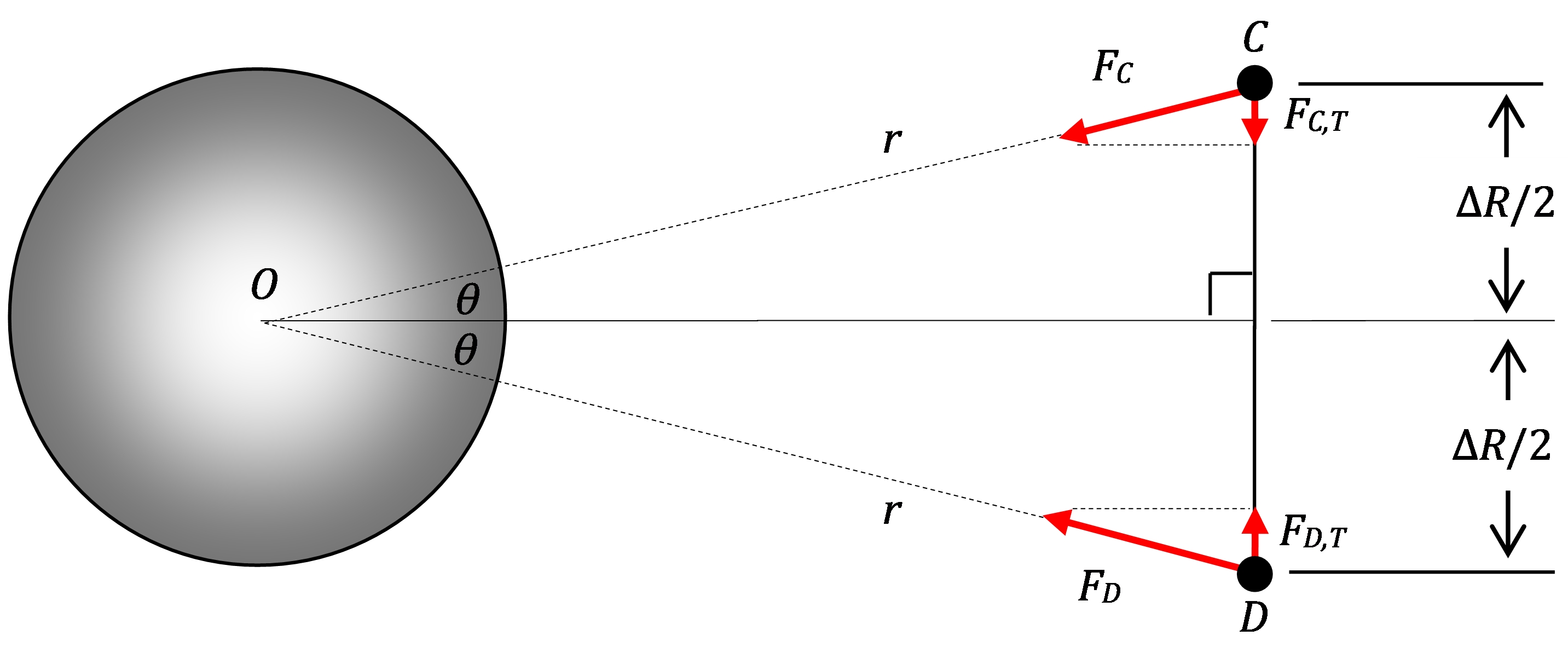}
  \caption{Transverse tidal forces on \textit{C} and \textit{D}, where $F_{C,T} = -F_{D,T}$.}
\end{figure}

From Eqs. (6) and (9), we see that $\Delta F_{T} = \Delta F_{R}/2$, meaning that if we know the radial tidal force, we automatically know the transverse force. We can therefore simplify the analysis by considering only the radial tidal force, as defined in Eq. (6), which we can write (omitting the subscript) as:

\begin{equation} 
\Delta F = \frac{2GMm\Delta r}{r^{3}}.
\end{equation}

We can further simplify the analysis if we now focus on the effects of tidal forces on LR, a small probe robot of mass $1kg$ and height $1m$ that has undertaken a one-way journey towards a black hole (see Fig. 5). Under these conditions, taking $\Delta r = 1m$ and $m = 1kg$ in Eq. (10), we have:

\begin{equation} 
\Delta F(N) = \frac{2GM}{r^{3}}, 
\end{equation}

where $G$, $M$ and $r$ must be expressed in the international system of units (SI) so that $\Delta F$ is expressed in Newtons ($N$). We can use Eq. (11) to calculate the tidal forces on LR at the earth's surface. Since the radius of the Earth is $R_{\bigoplus} = 6.37 \times 10^{6} m$ and its mass is $M_{\bigoplus} = 5.97 \times 10^{24} kg$, according to Eq. (11) the tidal force between the head and feet of LR will be $\Delta F_{\bigoplus} \cong 3 \times 10^{-6} N$. This value will serve as a comparison parameter for some calculations presented in the next section.

\begin{figure}[h]
  \centering
    \includegraphics[width=0.2\textwidth]{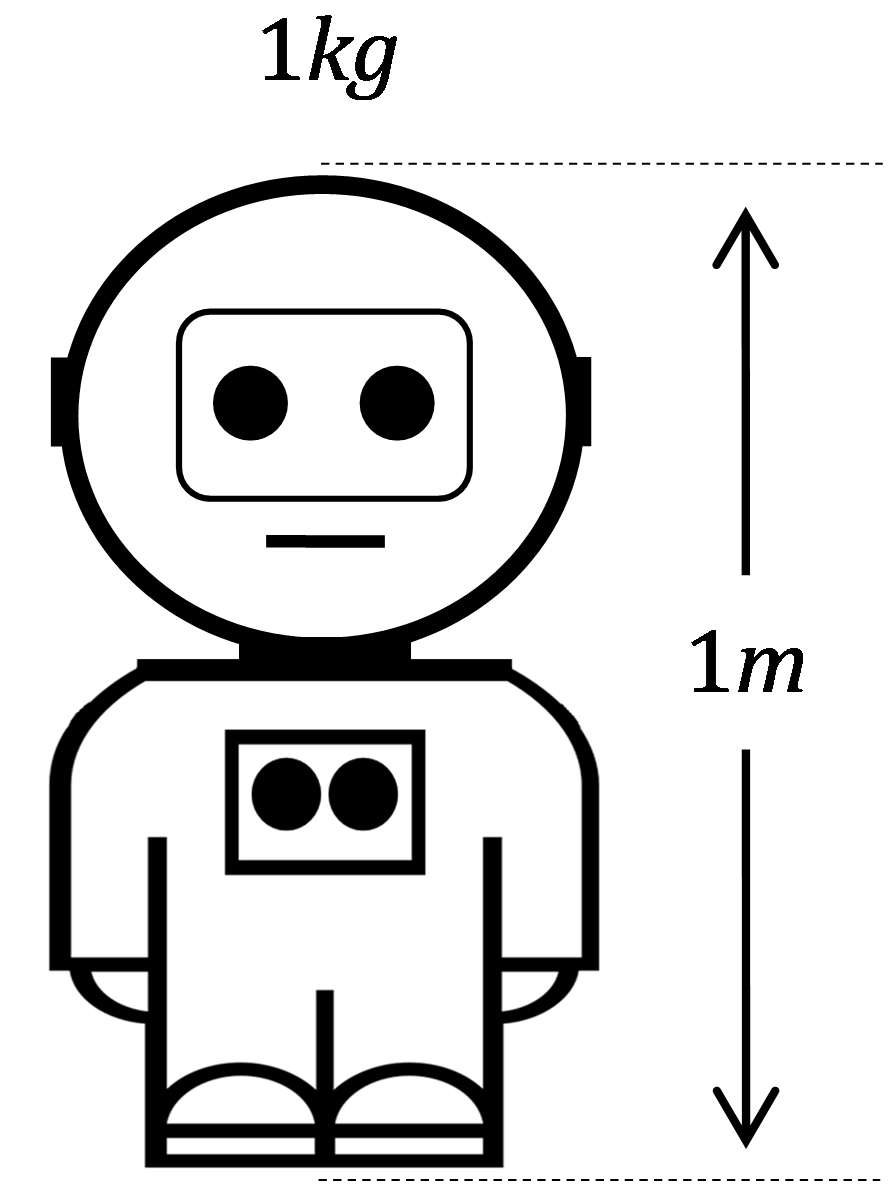}
  \caption{The Little Robot, LR.}
\end{figure}

\section{The little robot and spaghettification}
To describe the tidal forces exerted by a black hole on LR, it is convenient to rewrite Eq. (11) as a function of the Schwarzschild radius, Eq. (1):

\begin{equation} 
\Delta F(N) = \left( \frac{2GM}{c^{2}} \right) \frac{c^{2}}{r^{3}} = \frac{R_{S} c^{2}}{r^{3}}.
\end{equation}

We can measure the distance in the denominator in units of the Schwarzschild radius, $r/R_{S}$:

\begin{equation} 
\Delta F(N) = \frac{(c/R_{S})^{2}}{(r/R_{S})^{3}}.
\end{equation}

This equation gives the tidal forces in the outer region of a black hole ($r \geq R_{S}$). In relation to this point it is important to emphasize that although, basically, Eqs. (12) and (13) coincide with the results provided by general relativity, a rigorous mathematical treatment of this problem requires Einstein's theory. Newtonian physics only provides a first approximation whose main virtue is its simplicity, which makes it especially useful for pedagogical purposes.

To begin our analysis, we take $r/R_{S} = 1$ to obtain the tidal forces on LR at the horizon:

\begin{equation} 
\Delta F(N) = \frac{c^{2}}{R_{S}^{2}} = \frac{c^{6}}{4G^{2} M^{2}}.
\end{equation}

This equation gives us a first result of great importance, and which is counterintuitive: the tidal forces can be very small if a black hole is sufficiently large and massive. This result is a direct consequence of the fact that the Schwarzschild radius is proportional to the mass of the black hole, which is a fundamental prediction of general relativity. Although it is difficult to intuitively explain Eq. (14), we can briefly mention that tidal forces are the Newtonian equivalent of the curvature of space-time, which is the central concept of general relativity. Then, Eq. (14) expresses the fact that the curvature in the horizon depends inversely on the mass: the larger and more massive a black hole is, the smaller the curvature in the horizon.\\      

The largest that have been observed are \textit{supermassive black holes} (\textit{SM}), and the evidence strongly suggests that most galaxies have one of these objects at their centre\footnote{At the centre of our galaxy, there is a supermassive black hole known as \textit{Sagittarius} $A^{\ast}$, whose mass is estimated at $4 \times 10^{6} M_{\odot}$}. The mass of a supermassive hole can reach values of the order of $10^{10} M_{\odot}$, and according to Eq. (2), this implies that $R_{S} = 3 \times 10^{13} m$. Introducing this figure into the numerator of Eq. (13), and taking $c = 3 \times 10^{8 } m\cdot s^{-1}$, we obtain an upper bound for the tidal forces on LR caused by a supermassive black hole:

\begin{figure}[h]
  \centering
    \includegraphics[width=0.7\textwidth]{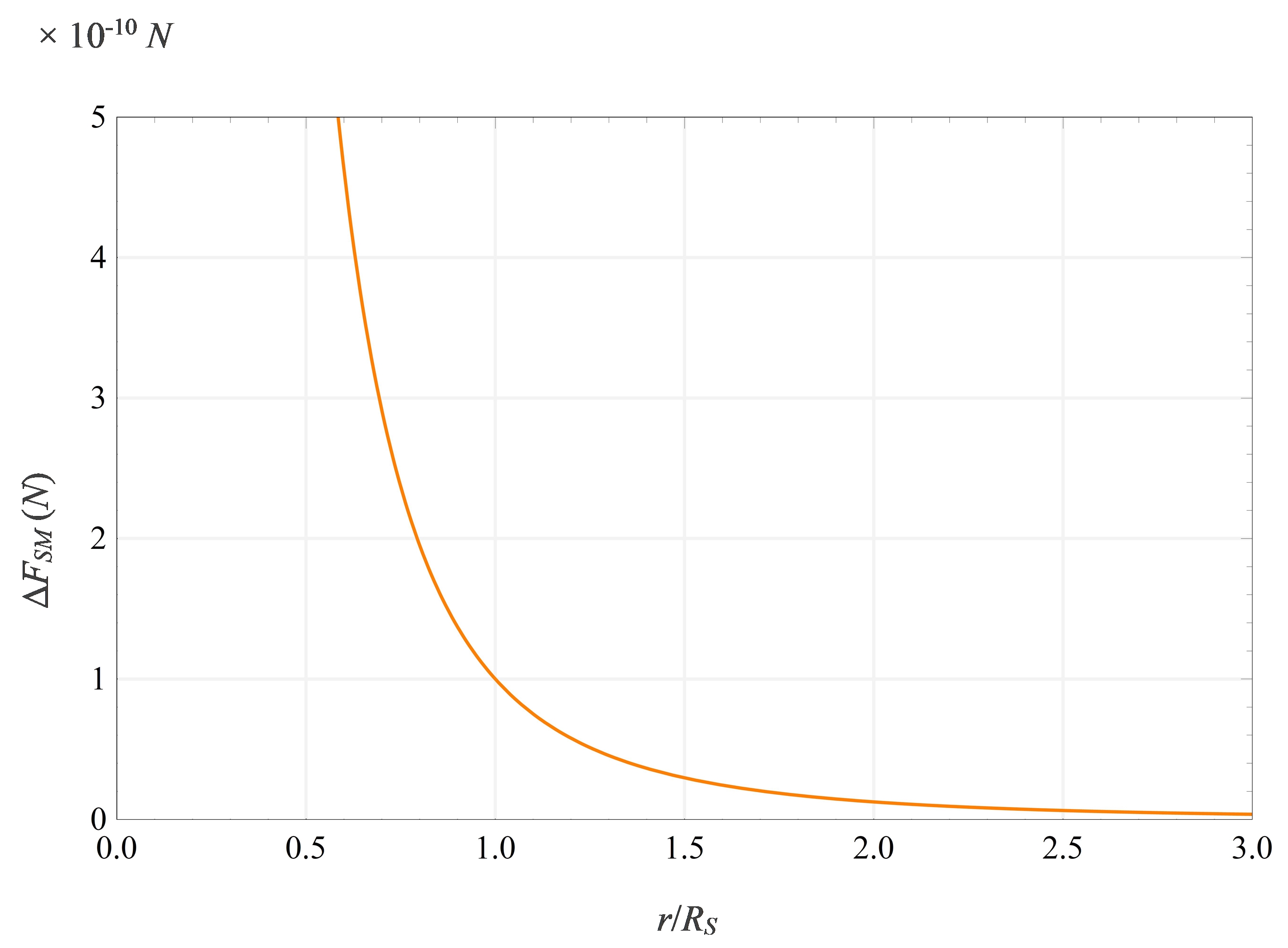}
  \caption{Plot of $\Delta F_{SM}$ (in units of $10^{-10} N$) versus distance (in units of $r/R_{S}$).}
\end{figure}

\begin{equation} 
\Delta F_{SM} (N) = \frac{10^{-10}}{(r/R_{S})^{3}}.
\end{equation}

Fig. 6 shows a graph of this expression. The horizontal axis is in units of $r/R_{S}$ and the vertical axis is in powers of $10^{-10} N$. It can be observed that $\Delta F_{SM}$ decreases rapidly with increasing $r$. Eq. (15) reveals that at the horizon, where $r/R_{S} = 1$, the tidal forces are on the order of $10^{-10} N$. This is a very small figure, and is lower by a factor of $10^{-4}$ than the tidal force on the earth's surface. Hence, supermassive holes do not generate spaghetti, at least in the outer region of the horizon.\\

The conclusions above contrast strongly with what happens with \textit{stellar black holes} (\textit{ST}), which are the lightest for which there is astronomical evidence. The smallest mass that these objects can have is on the order of $M_{\odot}$, which according to Eq. (2) corresponds to $R_{S} = 3\times 10^{3} m$. Introducing this value into Eq. (13), we obtain a lower bound for the tidal forces on LR in the case of a stellar black hole as follows:

\begin{equation} 
\Delta F_{ST} (N) = \frac{10^{10}}{(r/R_{S}^{3})}.
\end{equation}

Fig. 7 shows a graph of Eq. (16), where once again the horizontal axis is in units of $r/R_{S}$, and the vertical axis is in powers of $10^{10} N$. Qualitatively, this graph is identical to the previous one, and only the scale of the vertical axis changes.\\

Taking $r/R_{S} = 1$ in Eq. (16) results in $\Delta F_{ST} = 10^{10} N$, which is a factor $10^{16}$ greater than the tidal force on the earth's surface. This is a colossal force that would tear the unfortunate LR apart, meaning that we cannot continue to assume that its height is $1m$. In fact, long before reaching the horizon, LR will be spaghettified. This seems to invalidate the equations we have used so far; however, these equations are still valid if we assume that they describe the force experienced by two parts of the spaghettified body of LR that are separated by $1m$. Another way to analyse this situation is to imagine that LR is teleported to the horizon, so that the value of $10^{10} N$ provides the instantaneous tidal force.

\begin{figure}[h]
  \centering
    \includegraphics[width=0.7\textwidth]{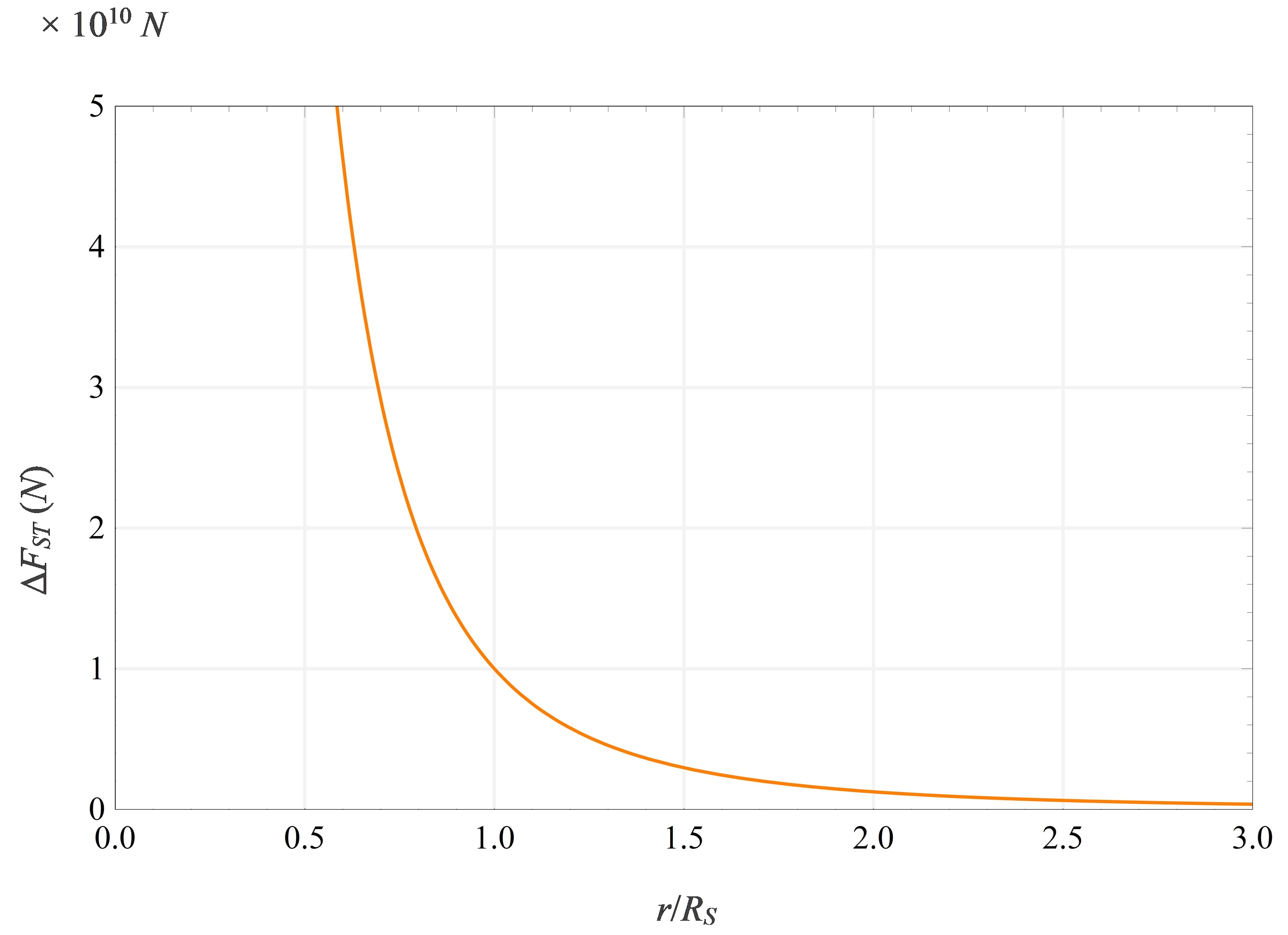}
  \caption{Plot of $\Delta F_{ST}$ (in units of $10^{10} N$) versus distance (in units of $r/R_{S}$).}
\end{figure}

If we opt for this last possibility, we can get an idea of what a tidal force of $10^{10} N$ means by imagining the situation illustrated in Fig. 8, where LR is at the earth's surface and is held rigidly by its head, while a $10^{9}kg$ object\footnote{The calculation used to arrive at this result is very simple: weight is defined as $mg$, where $g \cong 10 m\cdot s^{-2}$ at the earth's surface. Taking $mg \approx 10^{10} N$, we obtain $m \approx 10^{10} N / 10 m\cdot s^{-2} = 10^{9} kg$}  ($10^{6}$ tonnes) is hung on its feet. For this analogy to make sense, we must imagine that the figure shows the exact moment at which we hang the $10^{9}kg$ object from LR's feet, since an instant later, LR will be completely shattered. Also note that this analogy only illustrates the radial tidal force, although we know that there is also a compressive transverse force that is of the same order of magnitude as the radial stretching.\\

\begin{figure}[h]
  \centering
    \includegraphics[width=0.3\textwidth]{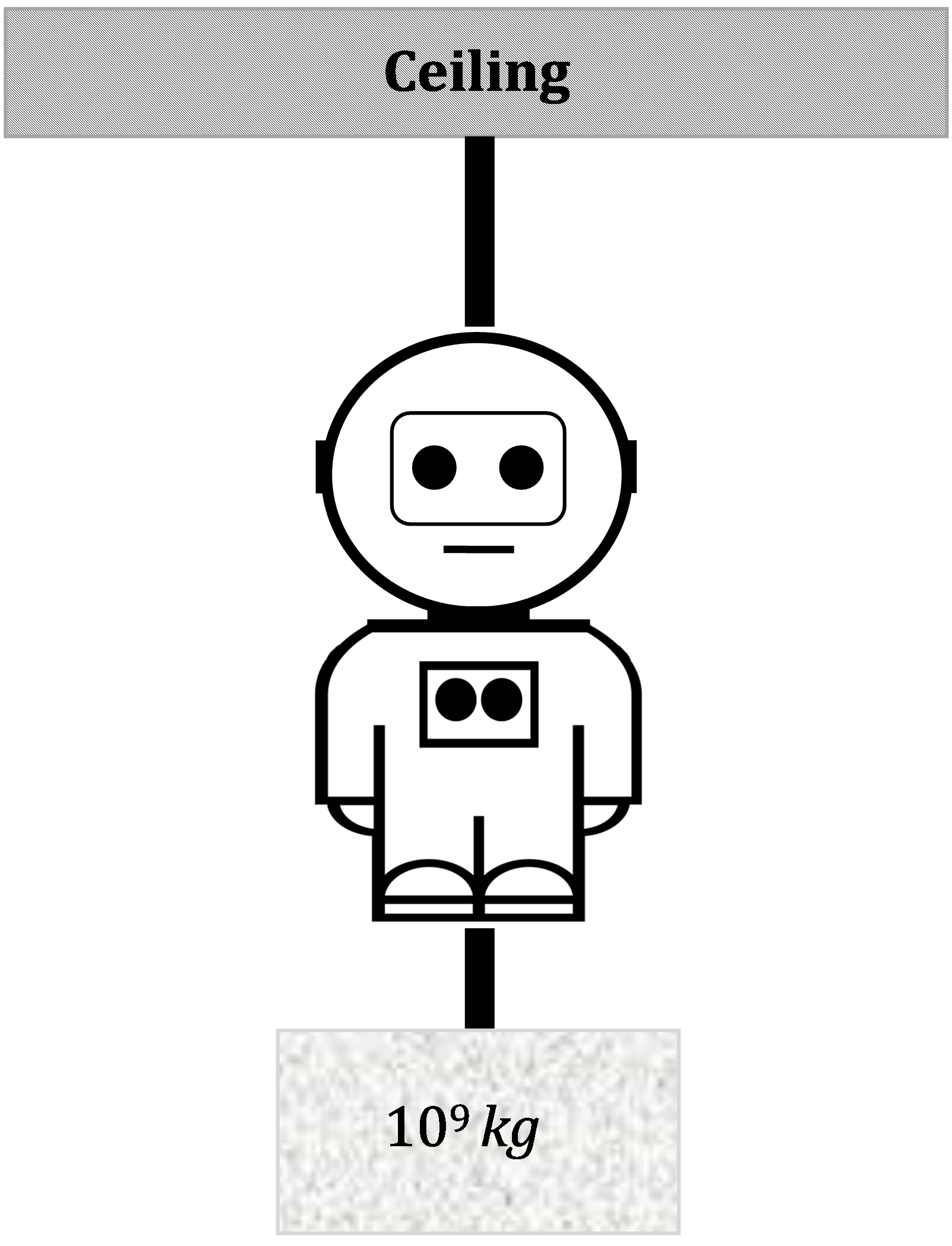}
  \caption{At the horizon of a stellar black hole, LR experiences an instantaneous radial stretching that is equivalent to being at the Earth's surface and held rigidly by its head while a $10^{9} kg$ object is dangled from its feet.}
\end{figure}

Finally, we can compare the tidal forces at the horizons of a supermassive black hole and a stellar hole. Taking $r/R_{S} = 1$ in Eqs. (15) and (16), we get $\Delta F_{ST} / \Delta F_{SM} = 10^{20}$, a result that confirms what we already know: tidal forces in stellar black holes are vastly more powerful than in supermassive holes.\\

\begin{figure}[h]
  \centering
    \includegraphics[width=0.6\textwidth]{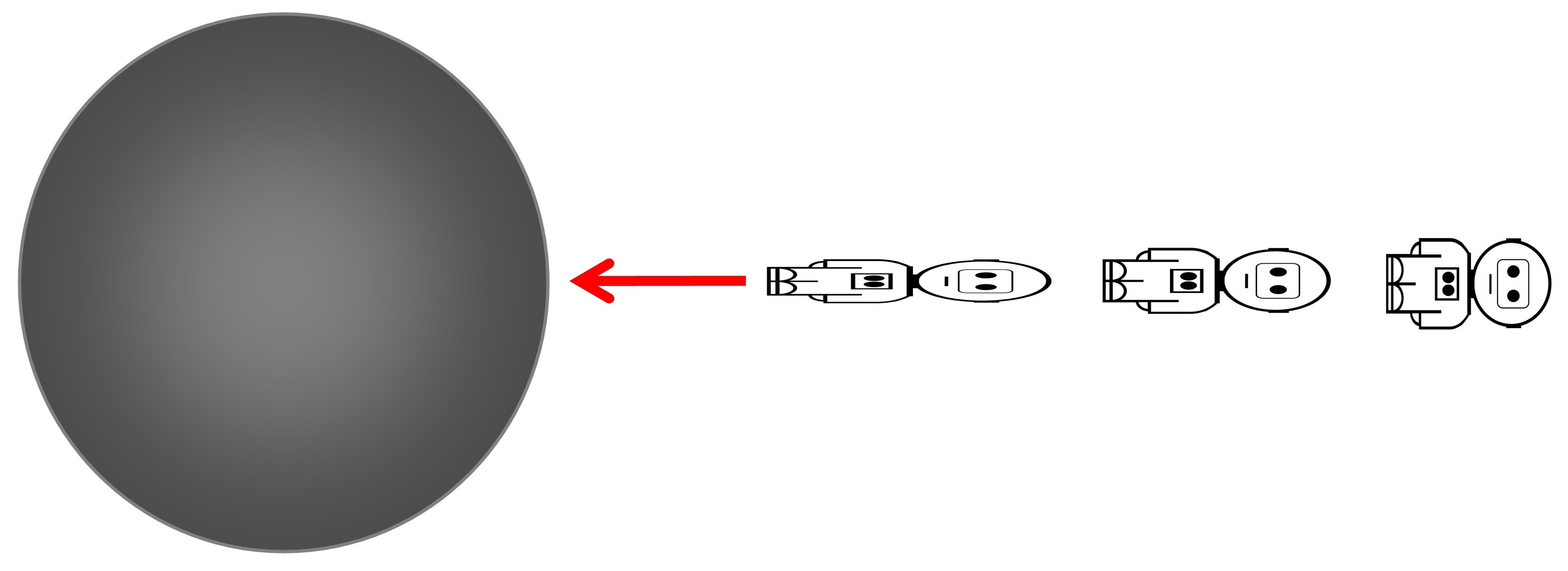}
  \caption{Three successive positions of LR as it falls freely into a black hole.}
\end{figure}

So far, we have only looked at what happens at the horizon. What happens to the tidal forces in the interior ($r < R_{S}$)? To answer this, we recall that a black hole contains a central singularity, which is a region of null size located at $r = 0$ at which all the mass is concentrated. An astronaut traversing the horizon of a supermassive black hole towards the interior would not at first feel anything, since $\Delta F$ is negligible, but the tidal forces would gradually increase and the astronaut would begin to spaghettified. This can be seen qualitatively from Eq. (12). Since the mass $M$ is concentrated in a central region of zero size, a decrease in $r$ can only produce an increase in $\Delta F$, and for $r \rightarrow 0$, $\Delta F \rightarrow \infty$. This conclusion also holds for a stellar mass hole, except that LR would be spaghettified long before reaching the horizon.\\

In short, as illustrated in Fig. 8, regardless of the mass or size of a black hole, spaghettification is inevitable for any object that gets close enough, and its effect will finally be so powerful that it will eventually will turn LR into a long, thin string of subatomic particles.\\

\section{Final comments: Can tidal forces be infinite?}
To find out how large tidal forces can become, we must determine whether there is a lower limit on the size of the matter confined at the centre of a black hole. Although general relativity implies that such matter occupies a null volume and has infinite density, meaning infinite tidal forces, there is a consensus that these conclusions are not valid. In other words, the mass-energy confined at the centre of a black hole must occupy a finite (albeit probably tiny) volume, and therefore must generate finite (albeit very large) tidal forces.\\

Does this mean that Einstein's theory is wrong? Like all physical theories, general relativity has a certain range of validity, and as we approach the centre of a black hole, where matter occupies a very small volume -perhaps smaller than the volume occupied by a subatomic particle- quantum effects become important, and relativistic physics is insufficient to provide a detailed description.\\

This reveals that, at their heart, black holes hide secrets that can only be revealed by unifying the laws of general relativity and quantum mechanics. Although a theory that is capable of achieving this unification, known as \textit{quantum gravity}, has not yet been elaborated in detail\footnote{The two main candidates for a theory of quantum gravity are \textit{Superstring Theory} and Q\textit{uantum Loop Gravity}.}, there are some ideas on which a certain consensus has been reached. For example, it is presumed that the degree of maximum concentration that matter can have is on the order of the \textit{Planck density} [6,7]:

\begin{equation} 
\rho _{P} \equiv \frac{m_{P}}{l_{P}^{3}} = \frac{c^{5}}{\hbar G^{2}} \sim 10^{96} kg\cdot m^{-3}.
\end{equation}

In this expression, $m_{P}$ and $l_{P}$ are the \textit{Planck mass} and the \textit{Planck length}, respectively, and are defined as [6]:

\begin{equation}
m_{P} \equiv \left( \frac{\hbar c}{G} \right)^{1/2} \sim 10^{-8} kg,\     l_{P} \equiv \left( \frac{G\hbar}{c^{3}} \right)^{1/2} \sim 10^{-35} m,    
\end{equation}

where $\hbar = h/2\pi = 1.05 \times 10^{-34} J\cdot s$ is the reduced Planck constant. Note that $l_{P}$, $\rho _{P}$ and $m_{P}$ bear the imprint of quantum gravity, since they combine the characteristic constants of general relativity ($c$ and $G$) and quantum mechanics ($\hbar$). Specialists consider $l_{P}$ to be the smallest unit of distance to which a physical meaning can be attributed. To appreciate how small $l_{P}$ is, we recall that the typical radius of an atomic nucleus is $\sim 10^{-15} m$, a factor $\sim 10^{20}$ larger than $l_{P}$.\\

If the above ideas are correct, instead of a central singularity of radius zero and infinite density, it is likely that at the centre of a black hole there is a region of radius $\sim l_{P}$ and density $\sim \rho _{P}$. As a result, the tidal forces near the centre should be finite, but very large. Exactly how large are these forces? Do they obey deterministic or probabilistic laws? What are the physical laws that govern them? For now, no one can answer these questions, but one thing is for sure: when an object is trapped by the gravity of a black hole, spaghettification and total destruction are inevitable.

\section*{Acknowledgments}
I would like to thank to Daniela Balieiro for their valuable comments in the writing of this paper. 

\section*{References}

[1] S.W. Hawking, A brief history of time, Bantam Books, New York, 1998.

\vspace{2mm}

[2] S.W. Hawking, The Universe in a Nutshell, Bantam Books, London, 2001.

\vspace{2mm}

[3] S. v Kontomaris, A. Malamou, A presentation of the black hole stretching effect, Physics Education. 53 (2018).

\vspace{2mm}

[4] J. Pinochet, Five misconceptions about black holes, Phys. Educ. 54 (2019) 55003.

\vspace{2mm}

[5] R.A. Serway, J.W. Jewett, Physics for Scientists and Engineers with Modern Physics, 8th ed., Thomson, Cengage Learning, 2010.

\vspace{2mm}

[6] J. Pinochet, The Hawking temperature, the uncertainty principle and quantum black holes, Phys. Educ. 53 (2018) 065004.

\vspace{2mm}

[7] R.J. Adler, Six easy roads to the Planck scale, American Journal of Physics. 78 (2010) 925–932.

\end{document}